\documentclass[twocolumn,aps,prl,showpacs,showkeys,superscriptaddress,notitlepage,longbibliography]{revtex4}
\usepackage[colorlinks=true,urlcolor=blue,citecolor=blue,linkcolor=blue]{hyperref}
\usepackage{graphicx}  
\usepackage{dcolumn}   
\usepackage{bm}        
\usepackage{CJKutf8}
\usepackage{amssymb}
\usepackage{amsfonts}
\usepackage{doi}
\usepackage{verbatim}
\usepackage{epstopdf}
\usepackage{lineno}
\usepackage{amsmath}
\usepackage{braket}
\usepackage{hyperref}

\begin{document}
\title{Transport Theory of Half-quantized Hall Conductance in a Semi-magnetic Topological Insulator}
\author{Humian Zhou}
\affiliation{International Center for Quantum Materials, School of Physics, Peking University, Beijing 100871, China}
\author{Hailong Li}
\affiliation{International Center for Quantum Materials, School of Physics, Peking University, Beijing 100871, China}
\author{Dong-Hui Xu}
\affiliation{Department of Physics, Chongqing University, Chongqing 400044, China}
\author{Chui-Zhen Chen}
\email{czchen@suda.edu.cn}
\affiliation{School of Physical Science and Technology, Soochow University, Suzhou 215006, China}
\affiliation{Institute for Advanced Study, Soochow University, Suzhou 215006, China}
\author{Qing-Feng Sun}
\affiliation{International Center for Quantum Materials, School of Physics, Peking University, Beijing 100871, China}
\affiliation{Collaborative Innovation Center of Quantum Matter, Beijing 100871, China}
\affiliation{CAS Center for Excellence in Topological Quantum Computation, University of Chinese Academy of Sciences, Beijing 100190, China}
\author{X. C. Xie}
\email{xcxie@pku.edu.cn}
\affiliation{International Center for Quantum Materials, School of Physics, Peking University, Beijing 100871, China}
\affiliation{Collaborative Innovation Center of Quantum Matter, Beijing 100871, China}
\affiliation{CAS Center for Excellence in Topological Quantum Computation, University of Chinese Academy of Sciences, Beijing 100190, China}
\begin{abstract}
Recently, a half-quantized Hall conductance (HQHC) plateau is experimentally observed in a semi-magnetic topological insulator heterostructure.
However, the heterostructure is metallic with a nonzero longitudinal conductance, which contradicts the common belief that quantized Hall conductance is usually observed in insulators.
In this work, we systematically study the surface transport of the semi-magnetic topological insulator with both gapped and gapless Dirac surfaces
in the presence of dephasing process.
In particular, we reveal that  the HQHC is directly related to the half-quantized chiral current along the edge of a strongly dephasing metal. The Hall conductance keeps a half-quantized value for large dephasing strengths,
while the longitudinal conductance varies with Fermi energies and dephasing strengths.
Furthermore, we evaluate both the conductance and resistance as a function of the temperature, which is consistent with the experimental results.
Our results not only provide the microscopic transport mechanism of the HQHC, but also are instructive for the probe of the HQHC in future experiments.

\end{abstract}

\maketitle

{\emph{Introduction.}}---
The  investigation of the Dirac fermions is an important paradigm in modern condensed matter physics \cite{Niemi1983,Jackiw1984,Semenoff1984,Fradkin1986,Haldane1988,Fu2007PRL,Fu2007PRB,Neto2009,Qi2011}.
Fundamentally, a single two-dimensional (2D) massive Dirac cone exhibits a hallmark half-quantized Hall conductance (HQHC) \cite{Niemi1983,Jackiw1984,Semenoff1984}, 
which plays an essential role in formulating various topological effects \cite{Qi2008,Nagaosa2011,Valley2015}.
Typical examples of this include quantum anomalous Hall (QAH)  effect proposed by Haldane, topological magnetoelectric effect in axion insulators and topological valley currents in 2D materials \cite{Haldane1988,Qi2008,Nagaosa2011,Valley2015}.
So far, many theoretical and experimental efforts have been exerted to observe the HQHC, but the direct observation  of the HQHC  remains elusive \cite{shen2011,K2014,yoshimi2015quantum,Novoselov2005,ZhangNat2005,Mogi2017,Chang2018,Lu2021}.
First, that's because the Dirac fermions of opposite chirality always appear in pairs in realistic systems according to fermion doubling theory \cite{Nielsen1981}.
More importantly, unlike the integer QAH effect that supports dissipationless chiral edge modes,
a single 2D massive Dirac cone does not possess topologically protected edge states and thus the mechanism of the HQHC goes beyond conventional paradigm of quantized transport.

\begin{figure}[bht]
	\centering
	\includegraphics[width=3.3in]{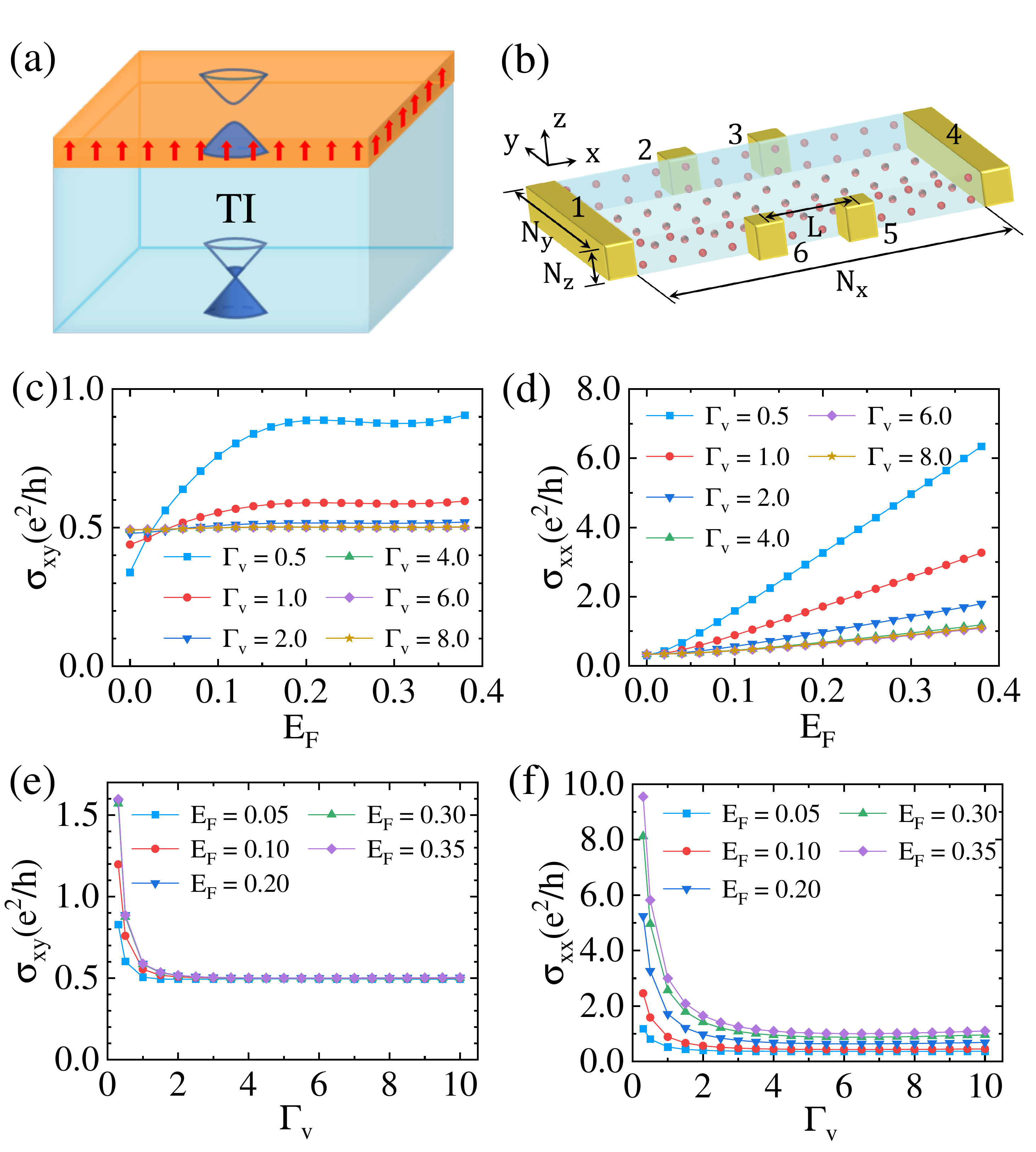}
	\caption{(Color online). (a) Schematic plot of a semi-magnetic TI with a gapped Dirac cone on the top surface and a gapless Dirac cone on the bottom surface, where the red arrows indicate the magnetization on the top surface.
		(b) Schematic plot of a six-terminal Hall-bar device, where the red balls represent the sites attached virtual leads. The distance of two nearest balls is $4$.
		(c) and (d) illustrate the Hall conductance $\sigma_{xy}$ and the longitudinal conductance $\sigma_{xx}$ against Fermi energy $E_F$ for different dephasing strengths $\Gamma_v$, respectively.
		(e) and (f) illustrate $\sigma_{xy}$ and $\sigma_{xx}$ against $\Gamma_v$  for different $E_F$, respectively.
		Here, $N_x=640$, $N_y=153$, $N_z=5$, $n_x=160$, $n_y=39$, $n_z=2$, and the distance between lead 5 and 6 is $L=40$.
		The model parameters are fixed as $A=1.0$, $B=0.6$, $M_0=1.0$ and $M_z=0.4$.
		\label{fig1} }
\end{figure}

Very recently, a breakthrough is made to the  experimental observation of the HQHC in a semi-magnetic topological insulator (TI) heterostructure \cite{Mogi2021}.
The semi-magnetic TI heterostructure consists of a single massive
 Dirac cone on the top surface and a massless Dirac cone on the bottom surface [see Fig.~\ref{fig1}(a)]
and thus it manifests as a metal in the transport measurement. 
It is found that the Hall and longitudinal resistance that directly measured in the experiment are non-quantized.
Surprisingly, by converting the resistance into the conductance, the Hall conductance is half-quantized in such a metallic phase, where the longitudinal conductance is nonzero.
This is in stark contrast to the common belief that quantized Hall conductance is generally observed in an insulating phase such as the QAH effect \cite{Haldane1988}.
Until now, the microscopic mechanism for the  HQHC in such a metallic heterostructure is unclear.

In this Letter, we uncover that the HQHC  originates from a half-quantized chiral current propagating  along the edge of a strongly dephasing metal, which we called classical metal.
To be specific, we study the surface transport of the semi-magnetic TI in the presence of dephasing process by Landauer-B{\"u}ttiker formula.
It is found that the the Hall conductance $\sigma_{xy}$  arrives at a half-quantized plateau when dephasing strength exceeds a critical value, whereas the longitudinal conductance  $\sigma_{xx}$ is nonzero.
Further, we show that the difference of transmission coefficients $t_d$ along the edge of the sample is half-quantized and independent of dephasing strength,  giving rise to a robust half-quantized chiral edge current.
Then, by establishing an analytic relation between $\sigma_{xy}$ and $t_d$,  we reveal that the half-quantized $t_d$ can lead to the HQHC in the classical metal limit, where the system size is much larger than the phase coherent length.
To compare with the experimental results \cite{Mogi2021}, we calculate the Hall and longitudinal conductance at low temperature. It is show that $\sigma_{xy}$ remains half-quantized regardless of temperature while  $\sigma_{xx}$ monotonically increases with the temperature, which is consistent with the experimental results.


{\em Model Hamiltonian and dephasing.}---We consider a semi-magnetic TI heterostructure shown in Fig.~\ref{fig1}(a).
The four-band effective Hamiltonian is $H=H_0 + H_M$, where
\begin{eqnarray}
  H_0({\bf k})&=& \sum_{i=x,y,z}Ak_i\sigma_x{\otimes}s_i+(M_0-Bk^2)\sigma_z{\otimes}s_0 \nonumber
\end{eqnarray}
describes the isotropic 3D TI \cite{Fu2007PRL}, with model parameters $A$, $B$, and $M_0$. $\sigma_i$ and $s_i$ are Pauli matrices for the orbital and spin degrees of freedom, respectively.
$H_M=M(z)\sigma_0{\otimes}s_z$ is the Zeeman splitting, where $M(z)$ takes the value $M_z$ on the top surface, and zero  elsewhere.
When $0<M_0<4B$, the 3D TI material is topologically nontrivial, and then it has a single gapless Dirac cone on each surface.
Due to the Zeeman splitting $H_M$,  an Dirac gap $\Delta=2M_z$  is opened up  on the top surface,
giving rise to a gapped Dirac cone with the HQHC.
Therefore, the semi-magnetic TI hosts a gapped Dirac cone and a gapless Dirac cone on the top and bottom surfaces, respectively [see Fig.~\ref{fig1}(a)].
Note that numerical results, calculated by using a real-space Kubo formula \cite{Prodan2009,Prodan2011},  demonstrate that the Hall conductance from the top surface
of the semi-magnetic TI is half-quantized when the Fermi energy $E_F$ is tuned the Dirac gap \cite{SM}. This coincides with the above analysis.

To investigate the surface transport of a six-terminal Hall-bar device [see Fig.~\ref{fig1}(b)], we discretize the Hamiltonian $H$ into $N_x \times N_y \times N_z$ cubic lattice sites.
The dephasing process is simulated by using  $n_x \times n_y$ and $n_x \times n_z$  B{\"u}ttiker's virtual leads on the bottom and side surfaces, respectively \cite{SM,Buttiker1986,Buttiker1988,Yanxia2008}.
According to Landauer-B{\"u}ttiker formula, the current in the lead $p$ can be expressed as:
\begin{eqnarray}\label{eq2}
	I_{p}=\frac{e^2}{h}\sum_{q\ne p}\left(T_{qp}V_{p}-T_{pq}V_{q}\right)
\end{eqnarray}
where $V_p$ is the voltage in the lead $p$.  $T_{pq}(E_F)=\mbox{Tr}[{\bm \Gamma}_p{\bm G}^r{\bm \Gamma}_q{\bm G}^a]$ is the transmission coefficient from the lead $q$ to the lead $p$, where the linewidth function ${\bm \Gamma}_p=i\left({\bm \Sigma}_p^r-{\bm \Sigma}_p^{r\dagger}\right)$ and the Green's function ${\bm G}^r=[{\bm G}^a]^{\dagger}=[E_F{\bm I}-H_{\mbox{cen}}-\sum_{p}{\bm \Sigma}_p^{r}]^{-1}$. ${\bm \Sigma}_p^r$ is the retarded self-energy due to the coupling to the lead $p$ and $H_{\mbox{cen}}$ is the lattice Hamiltonian of the semi-magnetic TI. For real leads ($p=1,2,...,6$), ${\bm \Sigma}_p^r=-i\frac{\Gamma_p}{2}{\bm I}_p$, where ${\bm I}_p$ is $4n_p\times 4n_p$ unit matrix and $n_p$ is the number of the sites coupling to the real lead p. For virtual leads, ${\bm \Sigma}_p^r=-i\frac{\Gamma_v}{2}$, where $\Gamma_v$ is the dephasing strength \cite{Yanxia2008}.
  When the longitudinal current  $I_x$ flows from lead $1$ to $4$, the Hall resistance $R_{xy}=(V_2-V_6)/I_x$ and longitudinal resistance $R_{xx}=(V_2-V_3)/I_x$ are obtained via Eq.~(\ref{eq2}).
Because the current only flow along the gapless surfaces (two side and one bottom surfaces), one can get  $\rho_{xy}=R_{xy}$, and $\rho_{xx}=R_{xx}/(L/W)$. Here, $L$ is the length between lead 2 and 3, and $W=(N_y+2(N_z-1)-1)$ is the total width of the gapless surfaces. The longitudinal and Hall conductance are obtained from the tensor relation: $\sigma_{xx}=\rho_{xx}/(\rho_{xx}^2+\rho_{xy}^2)$, and $\sigma_{xy}=\rho_{xy}/(\rho_{xx}^2+\rho_{xy}^2)$.

Now, we show the numerical results of the conductance below.  In Fig.~\ref{fig1}, the Hall conductance $\sigma_{xy}$ and the longitudinal conductance $\sigma_{xx}$ are both non-quantized for a small dephasing strength, e.g. $\Gamma_v=0.5$.
Remarkably, one finds that the Hall conductance $\sigma_{xy}$ decreases rapidly and arrived at a half-quantized plateau with  increasing $\Gamma_v$ [see Fig.~\ref{fig1}(e)],
while the longitudinal conductance $\sigma_{xx}$ remains nonzero [see Fig.~{\ref{fig1}} (f)]. We will explain it later.
After this half-quantization, $\sigma_{xy}$ is independent of Fermi energy $E_F$ [see Figs.~\ref{fig1}(c) and (e)].
In addition, $\sigma_{xx}$ increases almost linearly with $E_F$  in Fig.{\ref{fig1}}(d) because the  density of state of the gapless Dirac cone is proportional to $|E_F|$,
and $\sigma_{xx}$ decreases quickly with increasing $\Gamma_v$ in Fig.{\ref{fig1}} (f) due to the momentum relaxation introduced by the virtual leads \cite{Datta2007}.
\begin{figure}[bht]
	\centering
	\includegraphics[width=3.3in]{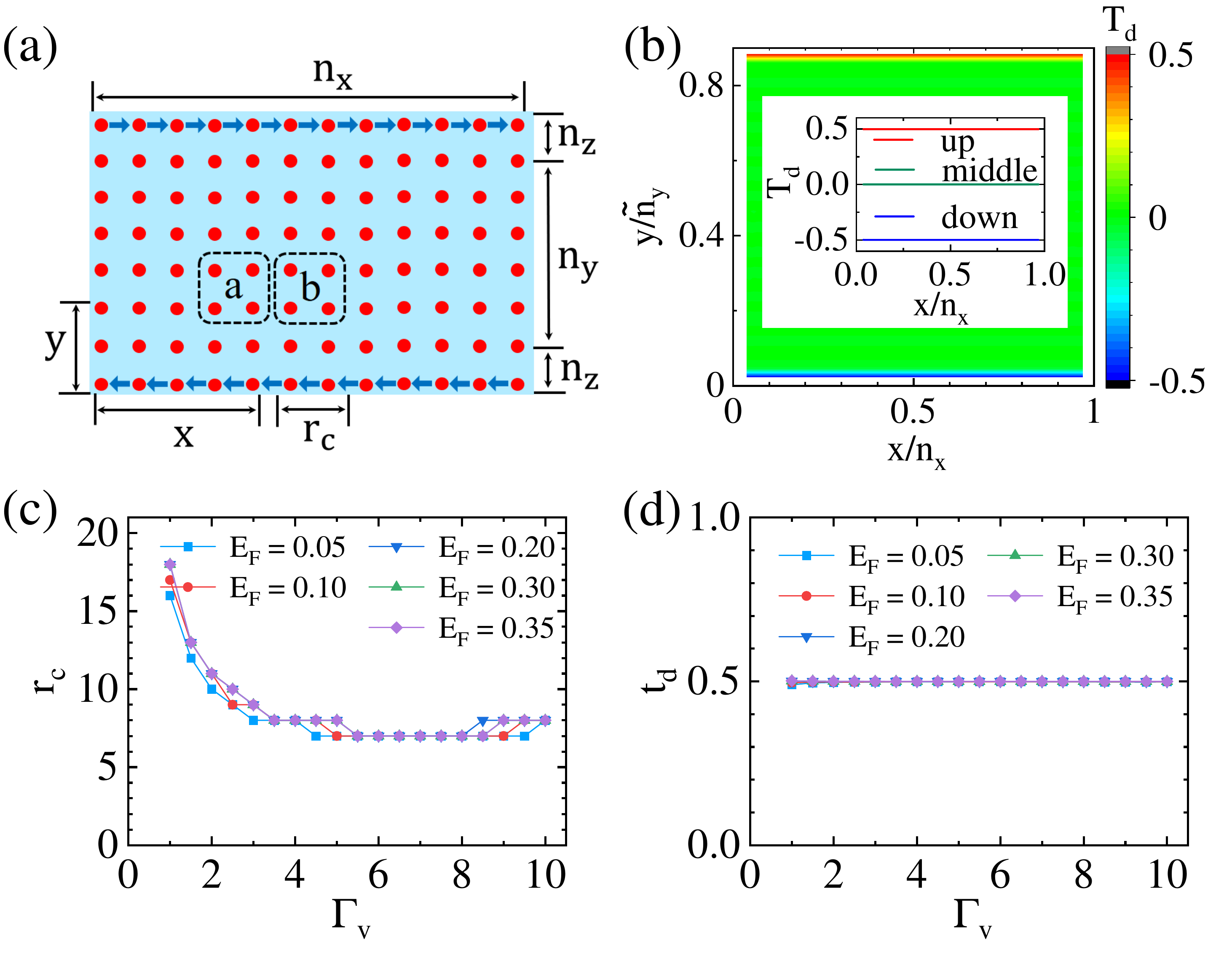}
	\caption{(Color online). (a) Lay the side and  bottom surfaces of the sample in Fig.~\ref{fig1} on the $xy$ plane. The red balls represent the sites attached to virtual leads and the blue arrows indicate chiral edge current.
	$a$ and $b$ are used to label the black box, and the size of black box is $r_c$. $(x,y)$ is the coordinate of the red ball in the lower right corner of the box $a$.
	(b) The space distribution of the different of transmission coefficients $T_d$ with $\Gamma_v=4.0$ and $E_F=0.3$. The inset show the value of $T_d$ in upper edge ($y=\widetilde{n}_y-r_c$), middle area ($y=(\widetilde{n}_y-r_c)/2$), and the lower edge ($y=1$). (c) and (d) show $r_c$ and $t_d$ as a function of $\Gamma_v$ for different $E_F$. Other parameters are the same as those in Fig.~\ref{fig1}.
		\label{fig2} }
\end{figure}

{\em Half-quantized chiral channel and HQHC.}--
 To get more insight into the origin of the HQHC, we investigate the transmission coefficients between the virtual leads.
 For convenience, we lay the side and bottom surfaces on the $xy$-plane [see Fig.~\ref{fig2}(a)].
 The red balls represent sites attached to the virtual leads. The total number of virtual leads in the $y$ direction is $\widetilde{n}_y=n_y+2(n_z-1)$.
 Numerical results show that the transmission coefficient $T_{pq}$ from lead $q$ to lead $p$ decays sharply with their distance $r_{pq}$ \cite{SM}.
 Thus, we can define a critical distance $r_c$: $T_{pq}=0.001T_1$ when $r_{pq}=r_c$, such that $T_{pq}$ can be neglected when $r_{pq}>r_c$.  Here, $T_1$ is the transmission coefficient between two nearest leads.
 In Fig.~\ref{fig2}(c), $r_c$ is proportional to the phase coherent length, and thus decreases with increasing the dephasing strength $\Gamma_v$.

In order to reveal the directionality of the transmission, we define the difference of transmission coefficients between black box $a$ and black box $b$ as $T_d(x,y)=T_{ba}-T_{ab}$, where $(x,y)$ is the spatial coordinate of the red ball in the lower right corner of the box $a$ [see Fig.~\ref{fig2}(a)].
Here  $T_{ba}=\sum_{p\in b,q\in a}T_{pq}$ the transmission coefficient from the box $a$ to the box $b$, and $p\in a$ means that lead $p$ is in box $a$ [see Fig.~\ref{fig2}(a)]. 
 Figure~\ref{fig2}(b) shows that $T_d=\pm 1/2$ in the upper edge and the lower edge, respectively, and zero elsewhere.
  Moreover, $t_d\equiv T_d(n_x/2,\widetilde{n}_y-r_c)$, the difference of transmission coefficient near the edges of top surface, keeps half-quantized for different $E_F$ and $\Gamma_v$ [see Fig.~\ref{fig2}(d)].
  The half-quantized $t_d$ means that there is  a half-quantized chiral channel
   or a half-quantized chiral current \cite{shen2011} on the edge of the top surface [see the blue arrows in Fig.~\ref{fig2}(a)], giving rise to the HQHC. The half-quantized chiral current is rather robust against the dephasing process, so we can obtain a perfect HQHC plateau for different dephasing strengths in Fig.~\ref{fig1}(a).

\begin{figure}[thb]
\centering
\includegraphics[width=3.3in]{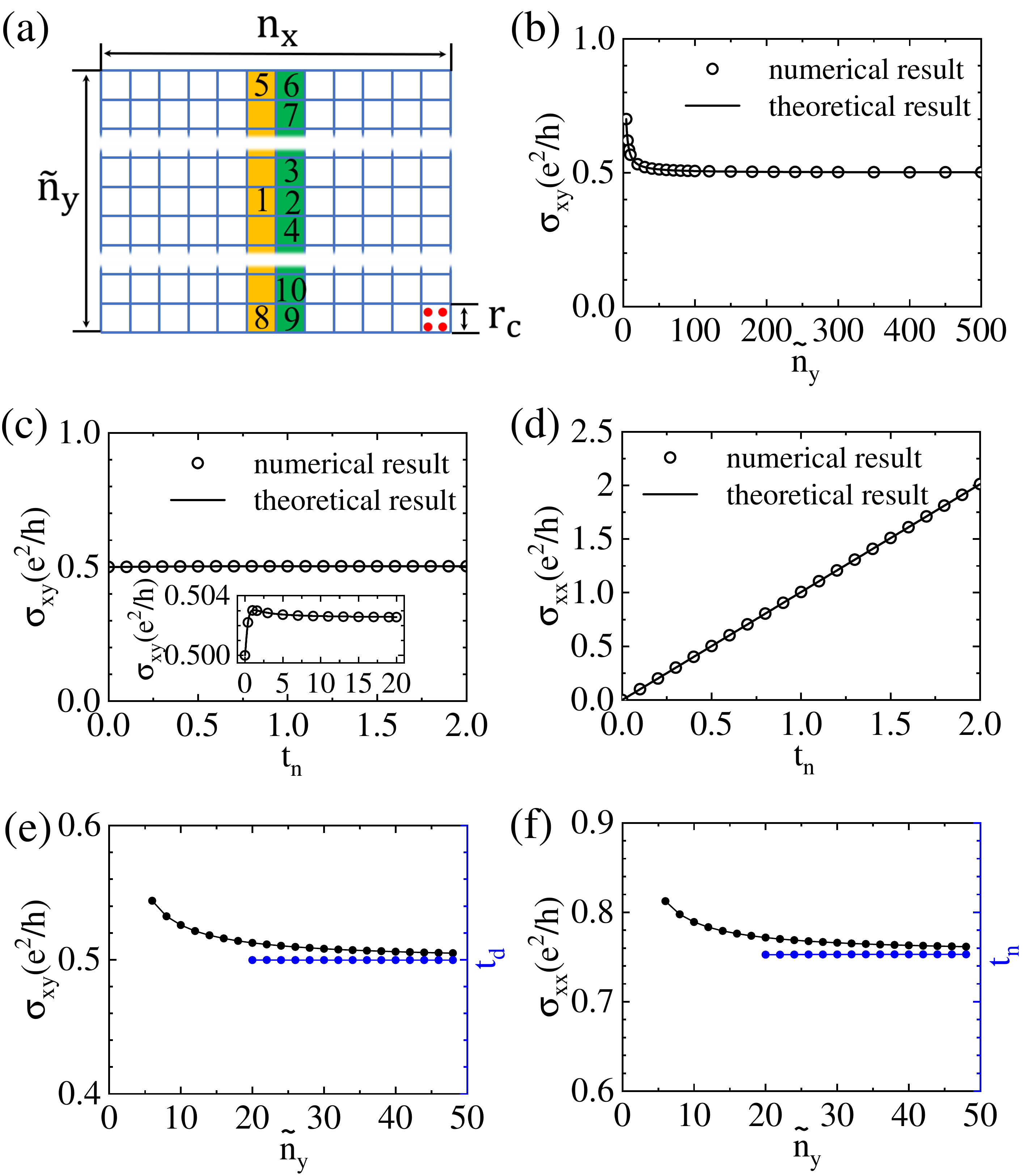}
\caption{(Color online). (a) Divide red balls in Fig.~\ref{fig2}.(a) by using blue boxes, and each blue box contain $r_c^2$ red balls. The numbers are used to label boxes. (b)-(d) Comparison of the theoretical results and numerical results for a simple case that $r_c=1$. (b) $\sigma_{xy}$ as a function of $\widetilde{n}_y$ with $t_n=1$. (c) and (d) $\sigma_{xy}$ and $\sigma_{xx}$ as a function of $t_n$ with $\widetilde{n}_y=200$. Other parameters in (b)-(d) are $n_x=7\widetilde{n}_y$ and $t_d=0.5$.
(e) $\sigma_{xy}$ and $t_d$ as a function of $\widetilde{n}_y$. (f) $\sigma_{xx}$ and $t_n$ as a function of $\widetilde{n}_y$. The parameters of (e) and (f): $E_F=0.2$, $\Gamma_v=3.0$, $n_z=2$, $n_y=\widetilde{n}_y-2(n_z-1)$, $n_x=4\widetilde{n}_y$, $N_z=5$, $N_y=4(n_y-1)+1$, $N_x=4n_x$, and other parameters are the same as those in Fig.~\ref{fig1}.
\label{fig3} }
\end{figure}

 To establish an analytic relation between $\sigma_{xy}$ and $t_d$,
we map the system in Fig.~\ref{fig2}(a) to a well-studied conductor-network model that describes conduction of a classical metal \cite{RNM1971,RNM1976}.
Here every $r_c^2$ red balls in Fig.~\ref{fig2}(a) are rearranged into one blue box in Fig~.\ref{fig3}(a).
Because $r_c$ much larger than the phase coherent length \cite{Yanxia2008}, the transmission between two adjacent blue boxes is incoherent, and the system can be regarded as the classical conductor-network model \cite{RNM1971,RNM1976}.
The conductance of conductor-network model in Fig~.\ref{fig3}(a) is determined by the total transmission coefficients between leads in two adjacent blue boxes, since $T_{pq}\ll 1$ when $r_{pq}>r_c$.
Using $I_{pq}=(T_{pq}V_q-T_{qp}V_p$) the current flowing from the lead $q$ to the lead $p$, 
one can get (see more details in Ref.~\cite{SM}):
\begin{eqnarray}\label{eq3}
\sigma_{xy}=\frac{e^2}{h}\left[1+\frac{r_c\alpha t_n+t_n^2}{t_d^2+t_n^2}\frac{1}{\widetilde{n}_y-1}
\right]t_d\nonumber\\
\sigma_{xx}=\frac{e^2}{h}\left[1+\frac{r_c\alpha t_n+t_n^2}{t_d^2+t_n^2}\frac{1}{\widetilde{n}_y-1}
\right]t_n
\end{eqnarray}
where $t_n=t_{12}+t_{13}+t_{14}$ and $\alpha=t_{56}+t_{57}+t_{89}+t_{810}-2t_n$, with $t_{ij}=\sum_{p\in j,q\in i}T_{pq}x_{pq}/r_c$. $x_{pq}=x_p-x_q$ and $x_{p}(x_{q})$ is the x-coordinate of the lead $p(q)$. $p\in i$ means that lead $p$ is in box $i$ [see Fig.~\ref{fig3}(a)].
Notably, in the large size limit when $\widetilde{n}_y\gg 1$, we have $\sigma_{xy}=t_d\frac{e^2}{h}$ and $\sigma_{xx}=t_n\frac{e^2}{h}$. This strongly demonstrates that the HQHC is directly related to the existence of half-quantized chiral channel $t_d$ on the edges of the top surface, while $\sigma_{xx}$ is contributed from the normal channels $t_n$.

Next, we demonstrate the validity of Eq.~(\ref{eq3}).
First,  we consider a simple model with $r_c=1$, where only the transmission coefficient between two nearest leads is non-zero (see more details in Ref.~\cite{SM}).
In Figs.~\ref{fig3}(b-d), we  evaluate $\sigma_{xy}$ and $\sigma_{xx}$  as a function of $t_n$ and $\tilde{n}_y$ numerically  by Eq.~(\ref{eq2}) and analytically by Eq.~(\ref{eq3}), respectively, when $t_d=0.5$.
The results are found to be fitting well which justifies Eq.(\ref{eq3}).
Furthermore, by starting from the realistic semi-magnetic TI Hamiltonian, we calculate $\sigma_{xy}$, $\sigma_{xx}$, $t_n$ and $t_d$ using Eq.~(\ref{eq2}) and plot them as a function $\tilde{n}_y$  in Fig.~\ref{fig3}(e-f).
It is found that $\sigma_{xy}$ ($\sigma_{xx}$) decreases and  converges to $t_d$ ($t_n$) when $\widetilde{n}_y$ increasing, which is coincident with the theoretical results of Eq.~(\ref{eq3})
and thus again confirm the validity of Eq.~(\ref{eq3}).
This strongly demonstrates the consistency and reliability of the obtained results.
In realistic systems, if $\widetilde{n}_y$ or $\Gamma_v$ is too small so that $\widetilde{n}_y/r_c\lesssim 1$, the half-quantized chiral current on the upper and lower edges will spatially mix due to finite $t_n$, which sabotages the quantization of $\sigma_{xy}$.
Therefore, we conclude that the dephasing process plays a key role in the separation of the two half-quantized chiral current in a metallic region, thus giving rise to the HQHC.
This mechanism of the HQHC is very different from that of the conventional quantized Hall conductance which is observed in an insulating phase \cite{Haldane1988}.


\begin{figure}[tbh]
\centering
\includegraphics[width=3.3in]{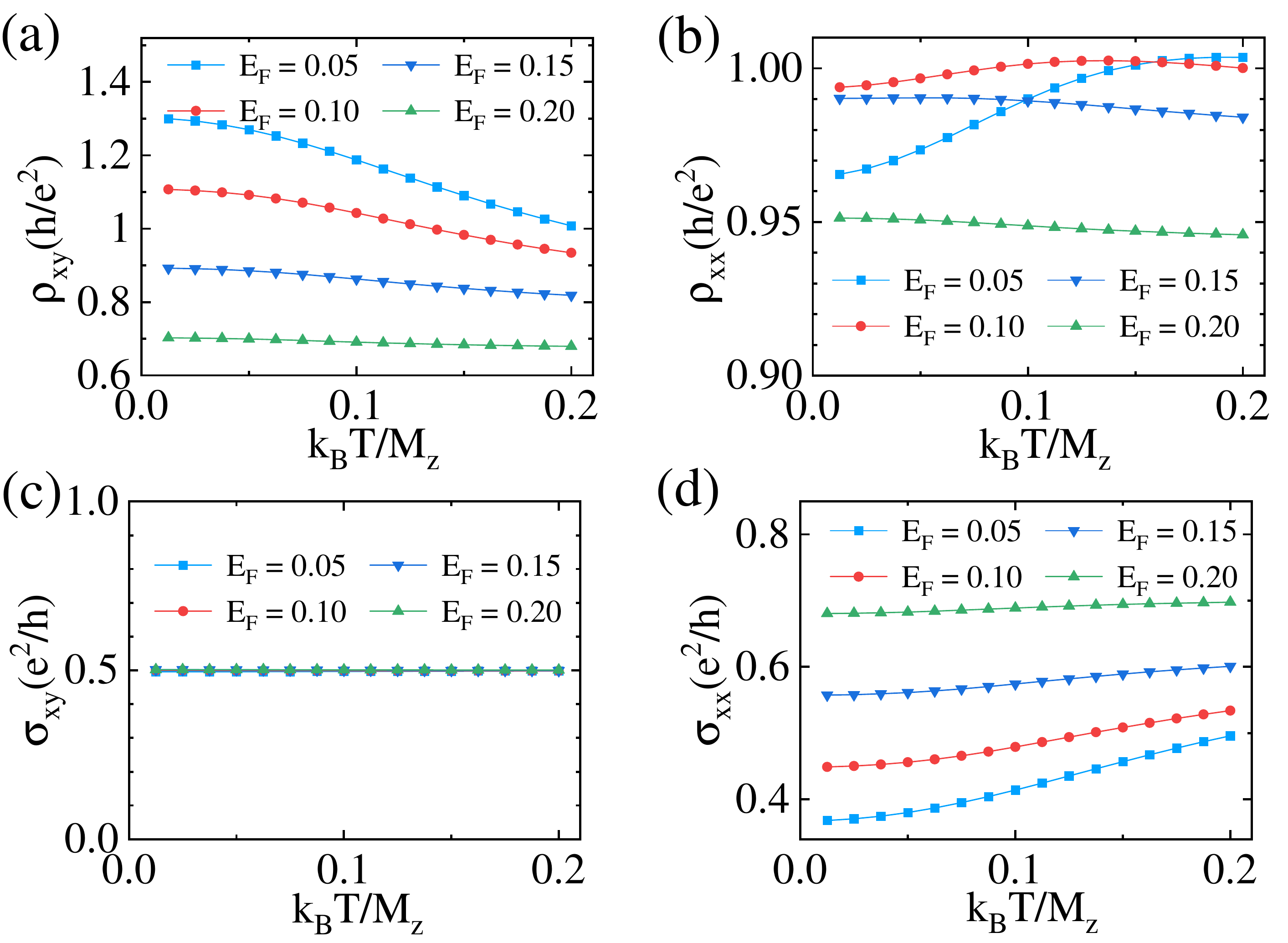}
\caption{(Color online). Hall resistance $\rho_{xy}$ (a),
	 longitudinal resistance $\rho_{xx}$ (b), Hall conductance $\sigma_{xy}$ (c) and
	 longitudinal conductance $\sigma_{xx}$ (d) as a function of temperature $T$. The curves correspond to different Fermi energy $E_F$. The dephasing strength $\Gamma_v=4.0$, and other parameters are the same as those in Fig.~\ref{fig1}.
\label{fig4} }
\end{figure}
{\em Comparison with experimental transport results}--
Recently, the Hall conductance $\sigma_{xy}$ is experimentally reported to be half-quantized in the semi-magnetic TI \cite{Mogi2021}. In contrast, the Hall resistance $\rho_{xy}$ and longitudinal resistance $\rho_{xx}$ that directly measured in the experiments, as well as the longitudinal conductance $\sigma_{xx}$, are all non-quantized.
Further, they show the temperature dependence of $\rho_{xx}$, $\rho_{xy}$, $\sigma_{xx}$ and $\sigma_{xy}$.
As a comparison, we calculate $\rho_{xx}$, $\rho_{xy}$, $\sigma_{xx}$ and $\sigma_{xy}$ as a function of $T$ by using Landauer-B{\"u}ttiker formula at non-zero temperature \cite{Data1995}: $I_{p}=\frac{e^2}{h}\sum_{q\ne p}\left(\widetilde{T}_{qp}V_{p}-\widetilde{T}_{pq}V_{q}\right)$. Here $\widetilde{T}_{pq}(T,E_F)=\int{T}_{pq}(E)(-\frac{\partial f_0}{\partial E})\text{d}E $ with the Fermi distribution $f_0=\left[e^{(E-E_F)/k_BT}+1\right]^{-1}$  and  the Boltzmann constant $k_B$.

In Figs.~\ref{fig4}(a-b), both $\rho_{xx}$ and  $\rho_{xy}$ are non-quantized and vary with $T$ and $E_F$.
Remarkably, when the resistance [in Figs.~\ref{fig4}(a-b)] is converted into the conductance,  $\sigma_{xy}$ is half-quantized  and $\sigma_{xx}$ is a monotonically increasing function of  $T$ for different $E_F$ [see Figs.~\ref{fig4}(c-d)], which is utterly consistent with the experimental results.
Based on the Landauer-B{\"u}ttiker formula at non-zero temperature, one gets $\sigma_{xy}(E_F,T)=\int{\sigma_{xy}}(E,T=0)(-\frac{\partial f_0}{\partial E})\text{d}E$ and $\sigma_{xx}(E_F,T)=\int{\sigma_{xx}}(E,T=0)(-\frac{\partial f_0}{\partial E})\text{d}E$ for $\tilde{n}_y \gg 1$ \cite{SM}.
This indicates that the dependence of conductance  on the Fermi energy  at zero temperature, determines the temperature dependence of conductance.
In Fig.~\ref{fig4}(d), $\sigma_{xx}$ increases with $T$ because $\sigma_{xx}$ increases almost linearly with $|E_F|$ at zero temperature,
while $\sigma_{xy}$ is independent of $T$ in Fig.~\ref{fig4}(c) because $\sigma_{xy}$ is independent of $E_F$ at zero temperature.
Then, $\rho_{xy}$ decreases with $T$ since $\rho_{xy}=\sigma_{xy}/(\sigma_{xx}^2+\sigma_{xy}^2)$ is a monotonically decreasing function of $\sigma_{xx}$.  $\rho_{xx}=\sigma_{xx}/(\sigma_{xx}^2+\sigma_{xy}^2)$ increases with $\sigma_{xx}$ for $\sigma_{xx}<\sigma_{xy}$ and decreases with $\sigma_{xx}$ for $\sigma_{xx}>\sigma_{xy}$. So $\rho_{xx}$ will increase first and then decrease with $T$ since $\sigma_{xx}<\sigma_{xy}$ at zero temperature for $E_F=0.05$ and 0.10, and monotonically decreases with $T$ since $\sigma_{xx}>\sigma_{xy}$ at zero temperature for $E_F=0.15$ and 0.20.

{\em Conclusion and discussion}-- In summary, we reveal that the half-quantized chiral current along the edge of a strongly dephasing metal is the origin of the HQHC of the semi-magnetic TI.
In reality, a 2D metal should be strongly influenced by the disorder, and the gapless Dirac cone could be driven into a critical metallic phase that can also host the HQHC \cite{Mogi2017,Chang2018}.
Nevertheless, the HQHC is measured in a hundreds-micron sample in the experiment, which far exceeds the dephasing length, and thus the system belongs to a classical metal.



{\emph{Acknowledgement.}}---  We thank M. Mogi, Jing-Yun Fang, Junjie Qi, Haiwen Liu and Hua Jiang for illuminating discussions.
This work was financially supported by National Key R and D Program of China (Grant No. 2017YFA0303301), NBRPC (Grant No. 2015CB921102), NSFC (Grants Nos. 11534001, 11822407, 11921005, 12074108, and 11704106), and also supported
by the Fundamental Research Funds for the Central Universities, the Strategic Priority Research Program of Chinese Academy of Sciences (DB28000000), and Beijing Municipal Science \& Technology
Commission (Grant No. Z191100007219013).
C.-Z.C. is also funded by the Priority Academic Program Development of Jiangsu Higher Education Institutions.

\bibliographystyle{apsrev4-1} 
\bibliography{ref}
\end{document}


\title{Supplementary Materials for ``Transport Theory of Half-quantized Hall Conductance in a Semi-magnetic Topological Insulator''}
\author{Humian Zhou}
\affiliation{International Center for Quantum Materials, School of Physics, Peking University, Beijing 100871, China}
\author{Hailong Li}
\affiliation{International Center for Quantum Materials, School of Physics, Peking University, Beijing 100871, China}
\author{Dong-Hui Xu}
\affiliation{Department of Physics, Chongqing University, Chongqing 400044, China}
\author{Chui-Zhen Chen}
\email{czchen@suda.edu.cn}
\affiliation{School of Physical Science and Technology, Soochow University, Suzhou 215006, China}
\affiliation{Institute for Advanced Study, Soochow University, Suzhou 215006, China}
\author{Qing-Feng Sun}
\affiliation{International Center for Quantum Materials, School of Physics, Peking University, Beijing 100871, China}
\affiliation{Collaborative Innovation Center of Quantum Matter, Beijing 100871, China}
\affiliation{CAS Center for Excellence in Topological Quantum Computation, University of Chinese Academy of Sciences, Beijing 100190, China}
\author{X. C. Xie}
\email{xcxie@pku.edu.cn}
\affiliation{International Center for Quantum Materials, School of Physics, Peking University, Beijing 100871, China}
\affiliation{Collaborative Innovation Center of Quantum Matter, Beijing 100871, China}
\affiliation{CAS Center for Excellence in Topological Quantum Computation, University of Chinese Academy of Sciences, Beijing 100190, China}
\date{\today }

\maketitle
\tableofcontents
\section{Layer-dependent Hall conductance of semi-magnetic topological insulator}
\begin{figure}[htbp]
	\includegraphics[scale=0.4]{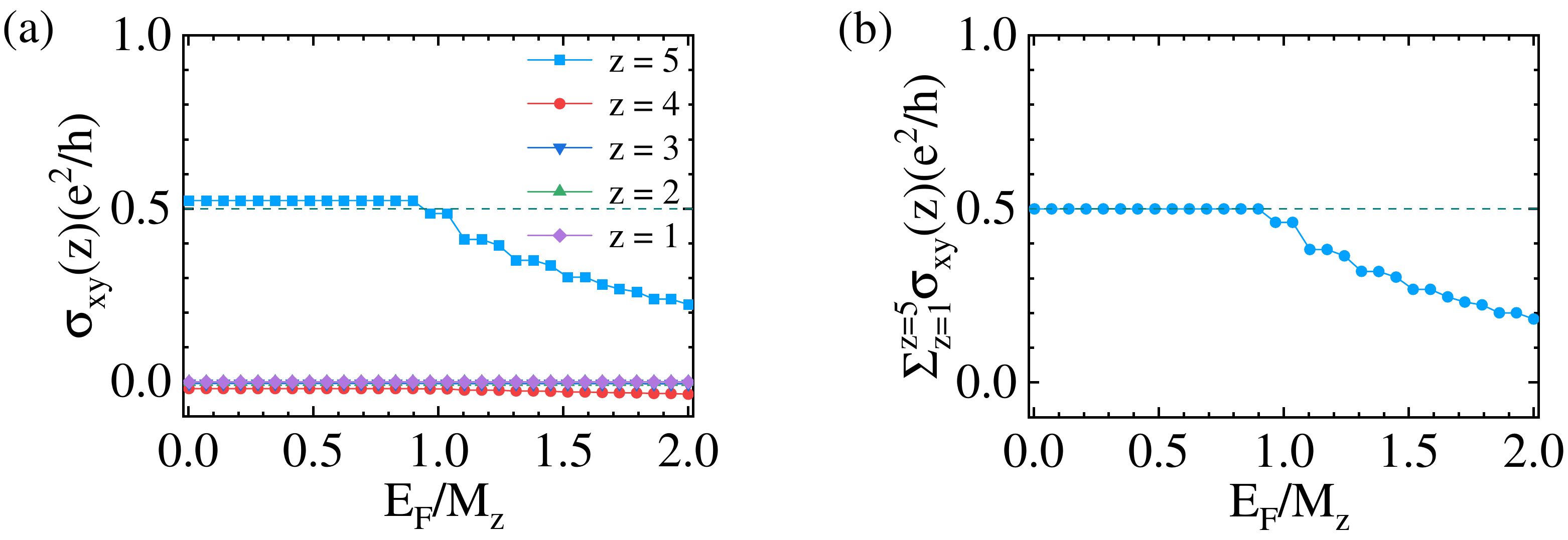}
	\caption{(a) The layer-dependent Hall conductance $\sigma_{xy}(z)$ as a function of Fermi energy $E_F$. $z$ is the layer index  with $z=1$ for the bottom layer and $z=5$ for the top layer. (b) The total Hall conductance as a fuction of $E_F$. Here, the size of sample is $N_x\times N_y\times N_z=40\times 40\times 5$. Other parameters are $A=1.0$, $B=0.6$, $M_0=1.0$ and $M_z=0.4$.
	}
	\label{figs1}
\end{figure}
In this section, we calculate the layer-dependent Hall conductance of semi-magnetic topological insulator (TI) by the real-space Kubo formula\cite{Prodan2009,Prodan2011}
\begin{eqnarray}
	\sigma_{xy}(z) = \frac{2\pi i e^2}{h} \mathrm{Tr}\left \{ P[-i[\hat{x},P],-i[\hat{y},P]]\right \}_z
	\label{eq1}
\end{eqnarray}
with periodic boundary condition in both the $x$ and $y$ direction.
$(\hat{x},\hat{y})$ denotes the position operator, and $\mathrm{Tr}\{...\}_z$ is trace over the wave functions of the $z$th layer. $P$ is the projector onto the occupied states of the  effective Hamiltonian $H$ of the semi-magnetic TI. Figure~\ref{figs1}(a-b) plots the layer-dependent Hall conductance $\sigma_{xy}(z)$ and the total conductance $\sum_{z=1}^{z=5}\sigma_{xy}(z)$ as a function of Fermi energy $E_F$, respectively. As $E_F$ is located in the Dirac gap, the total Hall conductance is half-quantized and the non-zero Hall conductance is mainly from the top surface.

\section{Lattice model Hamiltonian and dephasing}
We discretize the Hamiltonian $H$ in the main text on cubic lattices:
\begin{eqnarray}\label{eq11}
	H=&\left[
	\sum_{{\bf i}}{\bf c}_{\bf i}^{\dagger}{\mathcal M}_0{\bf c}_{\bf i}
	+\left(\sum_{{\bf i},j=x,y,z}{\bf c}_{\bf i}^{\dagger}{\mathcal T}_j{\bf c}_{{\bf i}+{\bf \hat{e}}_j}+\text{H.c.}\right)
	\right]
	+\sum_{{\bf i}\in \mbox{top surface}}{\bf c}_{\bf i}^{\dagger}{\mathcal M}_z{\bf c}_{\bf i}+H_d
\end{eqnarray}
where ${\mathcal M}_0=\left(M_0-6B/a^2\right)\sigma_z{\otimes}s_0$, ${\mathcal M}_z=M_z\sigma_0{\otimes}s_z$, and ${\mathcal T}_j=B/a^2\sigma_z{\otimes}s_0-iA/(2a)\sigma_x{\otimes}s_j$. Here, $a=1$ is the cubic lattice constant.
${\bf c}_{\bf i}$ (${\bf c}_{\bf i}^{\dagger}$) is the annihilation (creation) operator at site ${\bf i}$, and ${\bf \hat{e}}_j$ is a unit vector in the $j$ direction for $j=x,y,z$.
The first and the second terms in Eq. (\ref{eq11}) describe the semi-magnetic topological insulator (TI) in the central region. The last term $H_d=
\sum_{{\bf i},{\bf k}}\left[\epsilon_{{\bf k}}{\bf a}_{{\bf i},{\bf k}}^{\dagger}{\bf a}_{{\bf i},{\bf k}}+(t_{{\bf i},{\bf k}}{\bf a}_{{\bf i},{\bf k}}^{\dagger}{\bf c}_{{\bf i}}+\text{H.c.})\right]$ represents the Hamiltonian of virtual leads and real leads, and their couplings to the central sites.
 Note that we use the wide-band approximation for the real leads in the main text with parameters $\Gamma_1=\Gamma_4=0.9$, $n_1=n_4=N_z \times N_y$, $\Gamma_i=\Gamma_v$ and $n_i=n_z$ for $i=2,3,5,6$, where  $\Gamma_v$  is the dephasing strength.
Only electrons in the gapless surface states participate in the electrical transport when $E_F$ is located in the Dirac gap at zero or low temperature. Therefore, we just need to consider the dephasing process on the side and bottom surfaces.
We simulate the dephasing process by using $n_x\times n_y$ uniformly distributed B{\"u}ttiker's virtual leads on the bottom surface and $n_x\times n_z$ uniformly distributed B{\"u}ttiker's virtual leads on each side surface.
$N_i$ and $n_i$ are the number of total sites and selected sites in the $i$ direction for $i=x,y,z$, respectively.

In our calculations, real lead 1 and 4 act as current electrodes and other real leads act as voltage electrodes. Thus, the longitudinal current $I_1=-I_4\equiv I_x$ when a external bias is applied between lead 1 and lead 4 with $V_1=V/2$ and $V_4=-V/2$. Virtual leads act as dephasing, so the net current of each virtual lead is zero. 
Combining Landauer-B{\"u}ttiker formula\cite{Buttiker1988,Data1995} with those boundary conditions in the real and virtual leads, one can calculate the voltage of each real lead and the longitudinal current $I_x$. Then the Hall resistance $R_{xy}=(V_2-V_6)/I_x$ and longitudinal resistance $R_{xx}=(V_2-V_3)/I_x$ will be obtained.

\section{Derivation of the expressions of conductance at zero temperature}
\begin{figure}[htbp]
    \includegraphics[scale=0.4]{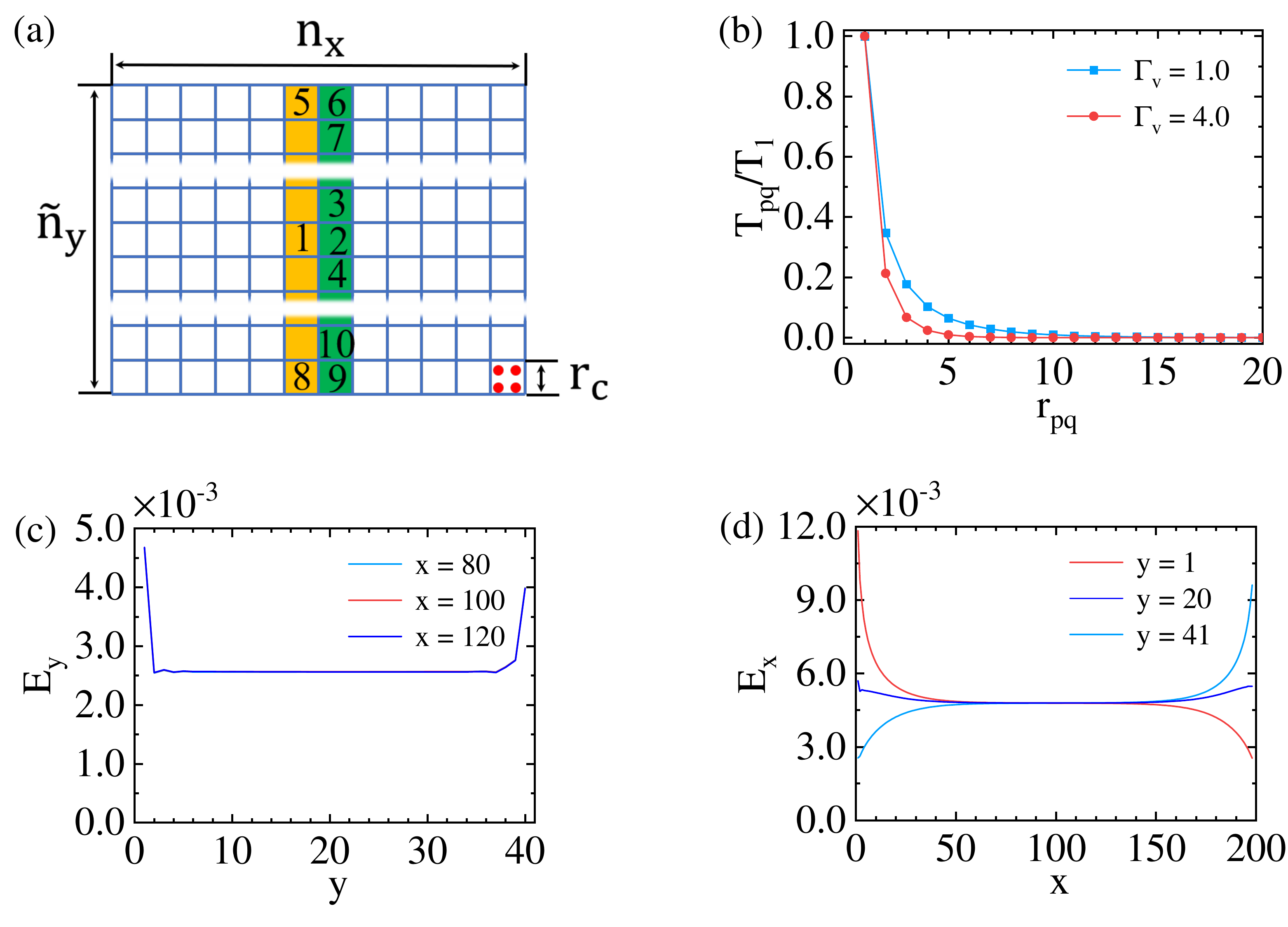}
    \caption{(color online) (a) Divide red balls in Fig.2.(a) in the main text by using blue boxes, and each blue box contain $r_c^2$ red balls. The numbers are used to label boxes.
    (b) $T_{pq}/T_1$ vs $r_{pq}$ for different $\Gamma_v$ with $E_F=0.3$. $T_{pq}$ is the transmission coefficient from virtual lead $q$ to virtual lead $p$, and $r_{pq}$ is the distance of those two leads.
     $T_1$ is the transmission coefficient between two nearest leads. (c) and (d) $E_x$ vs $x$ for different $y$
    and $E_y$ vs $y$ for different $x$. $(E_x,E_y)$ is the electric field and $(x,y)$ is the coordinate of red balls.
    $\Gamma_v=4.0$ and $E_F=0.3$ in (c) and (d). Here, $N_x=800$, $N_y=153$, $N_z=5$, $n_x=200$, $n_y=39$, $n_z=2$, and $\widetilde{n}_y=41$. Other parameters are $A=1.0$, $B=0.6$, $M_0=1.0$ and $M_z=0.4$.
    }
    \label{figs2}
  \end{figure}
In this section, we explains in detail how to derive the expressions of the Hall and longitudinal conductance. We lay the two side surfaces and the bottom surface of sample on the $xy$-plane, and divide red balls by blue boxes. Red balls represent sites attached virtual leads, so each blue box contains $r_c^2$ red balls [see Fig.~\ref{figs2}(a)]. Before derivation, let us discuss about three important assumptions and their validity.

 Firstly, we discuss about the critical length $r_c$. Fig.~\ref{figs2}(b) plot the transmission coefficient $T_{pq}$ from virtual lead $q$ to virtual lead $p$ versus their distance $r_{pq}$. $T_1$ is the transmission coefficient between two nearest leads. The results show that $T_{pq}$ decreases sharply with increasing $r_{pq}$. We define a critical distance $r_c$ satisfying that $T_{pq}=0.001T_1$ when $r_{pq}=r_c$. Thus $T_{pq}$ can be neglected when $r_{pq}>r_c$, which is the first assumption. 
 
 Secondly, we define longitudinal and transverse electric field as: $E_x\equiv V_q-V_{q+\hat{e}_x}>0$, $E_y\equiv V_{q+\hat{e}_y}-V_q>0$.Here, $V_q$ and $V_{q+\hat{e}_x}$ are the voltages of the lead $p$ and its nearest lead on the x direction, respectively. The numerical results demonstrate that the electric field is almost constant in the middle area of the gapless surface although it change when closing to the real lead 1 and lead 4. Therefore, the electric field of the green area and orange area can be thought to be constant, which is the second assumption. 
 
 The last assumption is that the difference of transmission coefficient between two nearest boxes is non-zero in the upper edge and lower edge, which is verified by numerical results of Fig.2(b) in the main text.
 
 Because  $T_{pq}$ is so small that that it can be neglected when $r_{pq}>r_c$, only the current $I_{pq}$ of two leads within one blue box or separately in two adjacent blue boxes is non-zero according to Landauer-B{\"u}ttiker formula \cite{Buttiker1988,Data1995}. $I_{pq}=e^2/h(T_{pq}V_q-T_{qp}V_p)$ is the net current flowing from lead $q$ to lead $p$.
 Then, one can get the total current flowing in the horizontal ($x$) direction:
 \begin{eqnarray}
 I_x &=& \sum_{p\in g,q\in o}I_{pq}
 =\sum_{p\in g,q\in o}(T_{pq}V_q-T_{qp}V_p) \nonumber \\
 &=&\sum_{p\in g,q\in o}T_{pq}(V_q-V_p)+\sum_{p\in g,q\in o}(T_{pq}-T_{qp})V_p \nonumber \\
 &=&\sum_{p\in g,q\in o}T_{pq}(x_{pq}E_x+y_{qp}E_y)+\sum_{p\in 6,q\in 5}(T_{pq}-T_{qp})V_p+\sum_{p\in 9,q\in 8}(T_{pq}-T_{qp})V_p \nonumber \\
 &=& (t_n\widetilde{n}_y+r_c\alpha)E_x+t_dV_u-t_dV_d \nonumber
\\ &=&t_nE_x\widetilde{n}_y+r_c\alpha E_x+t_dE_y(\widetilde{n}_y-1)
 \end{eqnarray}
Where the $V_u$($V_d$) is the voltage of the upper (lower) edge,  and $V_u-V_d=E_y(\widetilde{n}_y-1)$.  $p(q)\in g (o)$ means that lead $p(q)$ is in the green (orange) area [see Fig.~\ref{figs2}(a)].
Here, we have assumed $T_{pq}$ to be translational invariant on the gapless surface apart from its edges.
The first term of $I_x$ is the normal current, and the third term is the current from Hall effect.
The second term is the extra current from the difference between the conductance in the gapless surface and the conductance at its edge.
Here, $t_n=t_{12}+t_{13}+t_{14}$ and $\alpha=t_{56}+t_{57}+t_{89}+t_{810}-2t_n$. $t_{ij}=\sum_{p\in j,q\in i}T_{pq}x_{pq}/r_c$. $x_{pq}=x_p-x_q$ and $x_{p(q)}$ is the x-coordinate of the lead $p(q)$. $p\in i$ means that lead $p$ is in box $i$ [see Fig.~\ref{figs2}(a)].
And $t_n$ is isotropic and constant on the gapless surface (not include the edges of the surface).
Similarly, one can get the current flowing in the vertical ($y$)
direction $I_y=t_nE_yn_x+r_c\alpha E_y-t_dE_x(n_x-1)$. In our real model, $n_x\gg 1$ and $I_y=0$, thus $t_nE_y=t_dE_x$. Then $\rho_{xx}=E_x(\widetilde{n}_y-1)/I_x$ and $\rho_{xy}=E_y(\widetilde{n}_y-1)/I_x$. By converting the  resistance to the conductance, one can get:
\begin{eqnarray}\label{eq3}
	\sigma_{xy}(E_F)=\frac{e^2}{h}\left[1+\frac{r_c\alpha t_n+t_n^2}{t_d^2+t_n^2}\frac{1}{\widetilde{n}_y-1}
	\right]t_d
\end{eqnarray}
\begin{eqnarray}\label{eq4}
	\sigma_{xx}(E_F)=\frac{e^2}{h}\left[1+\frac{r_c\alpha t_n+t_n^2}{t_d^2+t_n^2}\frac{1}{\widetilde{n}_y-1}
	\right]t_n
\end{eqnarray}
\section{Transmission coefficients of the simple model}
\begin{figure}[htbp]
	\includegraphics[scale=0.3]{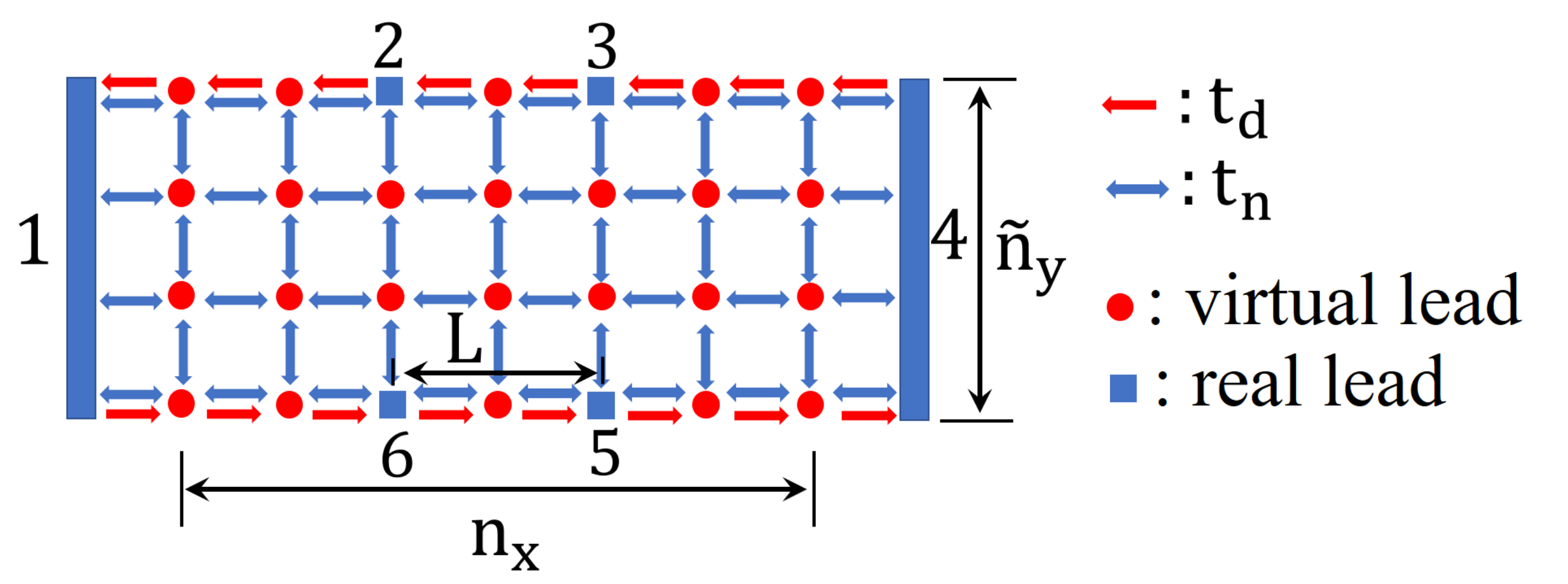}
	\caption{(color online) Schematic plot of the treansmission coefficients of the simple model. The blue square and rectangle represent the real lead while the red circle represents  the virtual lead. The double blue arrow indicate $T_{pq}=T_{qp}=t_n$ for two nearest leads. The single red arrow indicate an extra transmission $t_d$ coefficient from one lead to another lead that the red single arrow points to.
	}
	\label{figs3}
\end{figure}
In this section, we will illustrate the transmission coefficients of leads in the the simple model with $r_c=1$. For this simple model, only the transmission coefficient between two nearest leads is non-zero. As shown in Fig.~\ref{figs3},  $T_{p,q}=T_{q,p}=t_n$ for two nearest leads (include real leads and virtual leads). Considering the difference of transmission coefficients is non-zero in the edge, we have $T_{p+\hat{e}_x,p}=t_n$ and $T_{p,p+\hat{e}_x}=t_n+t_d$ in the upper edge, and $T_{p+\hat{e}_x,p}=t_n+t_d$ and $T_{p,p+\hat{e}_x}=t_n$ in the lower edge. Here, $p+\hat{e}_x$ represents the lead on the right-hand of lead $p$.
Obviously, $\alpha=t_d$.
The width of the sample is $W=\widetilde{n}_y-1$, and the distance between real lead 2 and lead 3 is $L=W$.
Given the transmission coefficients of leads, $\sigma_{xy}$ and $\sigma_{xx}$ can be calculated numerically by  Landauer-B{\"u}ttiker formula and then compared to Eqs.(\ref{eq3}-\ref{eq4}).

\section{Conductance at non-zero temperature}
In this section, we derive the Hall and longitudinal conductance at low temperature. In our numerical simulations, we use the Landauer-B{\"u}ttiker formula at non-zero temperature\cite{Data1995}:
\begin{eqnarray}
	 I_{p}=\frac{e^2}{h}\sum_{q\ne p}\left(\widetilde{T}_{qp}(T,E_F)V_{p}-\widetilde{T}_{pq}(T,E_F)V_{q}\right)
\end{eqnarray}
Here, $\widetilde{T}_{pq}(T,E_F)=\int{T}_{pq}(E)(-\frac{\partial f_0}{\partial E})\text{d}E $. $f_0=\left[e^{(E-E_F)/k_BT}+1\right]^{-1}$ is the Fermi distribution function, $E_F$ is the Fermi energy, and $k_B$ is the Boltzmann constant. In the section S3, we obtain the expression of conductance. Here, Similarly, one can get the total current flowing in the horizontal ($x$) direction $I_x=\widetilde{t}_nE_x\widetilde{n}_y+\widetilde{\alpha} E_x+\widetilde{t}_dE_y(\widetilde{n}_y-1)$ and the relation $\widetilde{t}_nE_y=\widetilde{t}_dE_x$ since the total current flowing in the vertical ($y$) direction $I_y=\widetilde{t}_nE_yn_x-\widetilde{t}_dE_x(n_x-1)=0$ with $n_x \gg 1$. Here, $\widetilde{t}_n(T,E_F)=\int{t}_{n}(E)(-\frac{\partial f_0}{\partial E})\text{d}E $, $\widetilde{t}_d(T,E_F)=\int t_{d}(E)(-\frac{\partial f_0}{\partial E})\text{d}E $ and $\widetilde{\alpha}(T,E_F)=\int{\alpha (E)r_c(E)(-\frac{\partial f_0}{\partial E})\text{d}E} $. Then, one gets:
\begin{eqnarray}\label{eq6}
	\sigma_{xy}(E_F,T)=\frac{e^2}{h}\left[1+\frac{\widetilde{\alpha} \widetilde{t}_n+(\widetilde{t}_n)^2}{(\widetilde{t}_d)^2+(\widetilde{t}_n)^2}\frac{1}{\widetilde{n}_y-1}
	\right]\widetilde{t}_d
\end{eqnarray}
\begin{eqnarray}\label{eq7}
	\sigma_{xx}(E_F,T)=\frac{e^2}{h}\left[1+\frac{\widetilde{\alpha} \widetilde{t}_n+(\widetilde{t}_n)^2}{(\widetilde{t}_d)^2+(\widetilde{t}_n)^2}\frac{1}{\widetilde{n}_y-1}
	\right]\widetilde{t}_n
\end{eqnarray}
When $n_y \gg 1$, $\sigma_{xy}(E_F,T)=\widetilde{t}_d$ and the expression in Eq.\ref{eq3} $\sigma_{xy}(E,T=0)=t_d(E)$. Then, one gets:
\begin{eqnarray}\label{eq8}
	\sigma_{xy}(E_F,T)=\int{\sigma_{xy}}(E,T=0)(-\frac{\partial f_0}{\partial E})\text{d}E
\end{eqnarray}
Similarly,
\begin{eqnarray}\label{eq9}
	\sigma_{xx}(E_F,T)=\int{\sigma_{xx}}(E,T=0)(-\frac{\partial f_0}{\partial E})\text{d}E
\end{eqnarray}
Here, $\sigma_{xy}(E,T=0)$ and $\sigma_{xx}(E,T=0)$ are the Hall and longitudinal conductance at zero temperature, respectively[see Eq.\ref{eq3} and \ref{eq4}]. Eq.\ref{eq8} and \ref{eq9} means that the total conductance comes from the carriers with energy near Fermi energy (that is,  $E_F-(\text{a few } k_BT)<E<E_F+(\text{a few }  k_BT)$).
Moreover, the dependence of conductance on the Fermi energy at zero temperature, determines the temperature dependence of conductance.

In the semi-magnetic TI, the Hall conductance at zero temperature $\sigma_{xy}(E,T=0)$ is independent of $E$ [see Fig.1(c) in the main text]. Then $\sigma_{xy}(E_F,T)=\sigma_{xy}(E_F,T=0)=0.5$ according to the Eq.\ref{eq8}, which is why $\sigma_{xy}(E_F,T)$ is half-quantized and independent of temperature in our numerical results and the experimental results. The longitudinal conductance $\sigma_{xx}\propto |E_F|$ due to the gapless Dirac cone[see Fig.1(d) in the main text]. Given $\sigma_{xx}(E,T=0)=c|E|$, and combine with the Eq.\ref{eq9}, one gets:
\begin{eqnarray}\label{eq10}
 \frac{\sigma_{xx}(E_F,T)}{\sigma_{xx}(E_F,T=0)}=1+\frac{2k_BT}{|E_F|}\mathrm{ln}(1+e^{-|E_F|/(k_BT)})
\end{eqnarray}
Obviously, $\sigma_{xx}(E_F,T)$ is a monotonically increasing function of  $T$, which coincides with our numerical results and the experimental results. Actually, only carriers with energy near Fermi energy (that is,  $E_F-(\text{a few } k_BT)<E<E_F+(\text{a few }  k_BT)$) have contributions to the longitudinal conductance. When $E_F>0$, holes with energy below Dirac point contribute to the extra conductance (the second term in Eq.\ref{eq10}), making the total conductance at low temperature larger than that at zero temperature. When the temperature is higher, or the Fermi energy is more close to the Dirac point, the number of holes participating in electric transport is more. Therefore, $\sigma_{xx}(E_F,T)$ is a monotonically increasing function of  $T$, and increases faster as the Fermi energy is more close to the Dirac point.

\bibliographystyle{apsrev4-1}
\bibliography{suppref.bib}